\documentclass[sigconf]{acmart}

\usepackage{booktabs} 
\usepackage{amsmath,amssymb,amsfonts}
\usepackage{algorithm,algorithmic}

\usepackage{graphicx}
\usepackage{subfig}
\usepackage{array}

\newcolumntype{P}[1]{>{\centering\arraybackslash}p{#1}}
\newcolumntype{M}[1]{>{\centering\arraybackslash}m{#1}}

\setcopyright{acmcopyright}

\hypersetup{draft}

\begin{document}
\title{Understanding the Socio-Economic Disruption in the United States during COVID-19's Early Days}


 \author{Swaroop Gowdra Shanthakumar, Anand Seetharam, Arti Ramesh}
 \affiliation{
   \institution{Computer Science Department, SUNY Binghamton}
}
 \email{ (sgowdra1, aseethar, artir)@binghamton.edu}


\begin{abstract}
In this paper, we collect and study Twitter communications to understand the socio-economic impact of COVID-19 in the United States during the early days of the pandemic. Our analysis reveals that COVID-19 gripped the nation during this time as is evidenced by the significant number of trending hashtags. With infections soaring rapidly, users took to Twitter asking people to self isolate and quarantine themselves. Users also demanded closure of schools, bars, and restaurants as well as lockdown of cities and states. The communications reveal the ensuing panic buying and the unavailability of some essential goods, in particular toilet paper. We also observe users express their frustration in their communications as the virus spread continued. We methodically collect a total of 530,206 tweets by identifying and tracking trending COVID-related hashtags. We then group the hashtags into six main categories, namely {\it 1)} General COVID, {\it 2)} Quarantine,  {\it 3)} Panic Buying, {\it 4)} School Closures, {\it 5)} Lockdowns, and {\it 6)} Frustration and Hope,  and study the temporal evolution of tweets in these hashtags. We conduct a linguistic analysis of words common to all the hashtag groups and specific to each hashtag group. Our preliminary study presents a succinct and aggregated picture of people's response to the pandemic and lays the groundwork for future fine-grained linguistic and behavioral analysis.
\end{abstract}
%
%

%
%


\maketitle


\section{Introduction}
\label{sec:intro}

COVID-19 (also known as the novel coronavirus) is the world's first global pandemic and has affected humans in all countries of the world. While humanity has seen numerous epidemics including a number of deadly ones over the last two decades (e.g., SARS, MERS, Ebola), the grief and disruption that COVID-19 will inflict is incomparable (and perhaps unimaginable). At the time of writing this paper, COVID-19 is still rapidly spreading around the world and projections for the next few months are grim and extremely disconcerting.

With no cure in sight and with the chances of COVID-19 reemerging for a second (or multiple) time(s) even after the world manages to contain this first outbreak, it is critical that we understand and analyze the socio-economic disruptions of the first outbreak, so that we are better prepared to handle it in the future. Additionally, with ever-increasing  mobility of humans and goods, it is only prudent to assume that such epidemics are likely to occur in the future. The learnings from COVID-19 will also enable humankind to prevent such epidemics from transforming into global pandemics and minimize the socio-economic disruption.

In this preliminary work, our goal is to analyze the socio-economic disruption caused by COVID-19 in the United States of America, understand the chain of events that occurred during the spread of the infection, and draw meaningful conclusions so that similar mistakes can be avoided in the future.  Though Twitter data has previously been shown to be biased \cite{morstatter:2017discovering}, Twitter has emerged as the primary media for people to express their opinion especially during this time and our study offers a perspective into the impact as self-disclosed by people in a form that is easily understandable and can be acted upon. We summarize our main contributions below.



\begin{itemize}
\item  We collect 530,206 tweets from Twitter between March $14^{th}$ to March $24^{th}$, a time period when the virus significantly spread in the US and quantitatively demonstrate the socio-economic disruption and distress experienced by the people. Calls for closures started off with schools (e.g., \#closenycschools), then moved on to bars and restaurants (e.g., \#barsshut), and finally to entire cities and states (e.g., \#lockdownusa). While these  calls were initially mainly confined to the Seattle, Bay Area, and  NY regions (e.g., \#seattleshutdown, \#shutdownnyc), they later expanded to include other parts of the country (e.g., \#shutdownflorida, \#vegasshutdown).  Alongside, panic buying and  hoarding escalated with essential items particularly toilet paper becoming unavailable in stores (e.g., \#panicbuying, \#toiletpapercrisis).

\item We  observe increased calls for social distancing, quarantining, and working from home to limit the spread of the disease (e.g., \#socialdistancingnow, \#workfromhome). To slow the exponential increase in the number of infections, people also rallied for flattening the curve and staying at home for extended periods (e.g., \#flattenthecurve). The challenges of working from home also surface in communications (e.g., \#stayhomechallenge). With the passage of time, we see an increased fluctuation in emotions with some people expressing their anger at individuals flouting social distancing calls (e.g., \#covidiots), while others rallying people to fight the disease (e.g., \#fightback) and to save workers (e.g., \#saveworkers).

\item We group the hashtags into six main categories, namely {\it 1)} General COVID, {\it 2)} Quarantine, {\it 3)} School Closures, {\it 4)}  Panic Buying, {\it 5)} Lockdowns,  and {\it 6)} Frustration and Hope,  to quantitatively and qualitatively understand the chain of events. We observe that general COVID  and quarantine related messages remain trending throughout the duration of our study. We observe calls for closing schools and universities peaking in the middle of March and then reducing when the closures go into effect. We observe a similar trend with panic buying. Lockdowns also have a significant number of tweets with calls initially being focused on closure of bars, followed by cities and then states. Tweets in the frustration and hope hashtag group have an overall increasing trend as the struggle with the virus mounts.

\item We then present a linguistic analysis of the tweets in the different hashtag groups and present the words that are representative of each group.  We observe that words such as \textit{family}, \textit{life}, \textit{health} and \textit{death} are common across hashtag groups. We  observe mentions to \textit{mental health}, a possible consequence of social isolation. We also observe solidarity for essential workers and gratitude towards them (\#saveworkers).

\end{itemize}
 Our preliminary study unearths and summarizes the critical public responses surrounding COVID-19, paving the way for more insightful fine-grained linguistic and graph analysis in the future.

\section{Data and Methods}
\label{sec:data}

In this section, we discuss our methodology for data collection from Twitter to  investigate the socio-economic distress and disruption  in the United States caused by COVID-19 during its early days. 
\subsection{Data Collection}
We collect data using the Twitter search API. The results presented in this paper are based on the data collected from March 14 to March 24, 2020. We track the trending COVID related hashtags every day and collect the tweets in those specific hashtags. We repeat this process to collect a total of 530,206 tweets during this time period.

\begin{table}[ht]
\caption{Number of Tweets by Category}
\vspace{-2 mm}
\begin{center}
\begin{tabular}{|c|c|}
\hline
\textbf{Category} & \textbf{Number of Tweets} \\
\hline
General Covid &  4,81,398 \\
 \hline
Quarantine &  142,297 \\
\hline
 Lockdowns & 14,709 \\

 \hline
 Frustration and Hope & 13,084 \\
 \hline
Panic Buying & 10,855 \\
 
 \hline
School Closures & 2,133 \\
  \hline
\end{tabular}
\label{tab_total_tweets}
\end{center}
\end{table}

\begin{table*}[ht]
\caption{Hashtags by Category}
\vspace{-2 mm}
\begin{center}
\begin{tabular}{|c|c|}
\hline
\textbf{Category} & \textbf{Hashtags} \\
\hline
General COVID  &  \#covid19, \#COVID19, \#Covid19, \#covid\_19,  \#covid, \#corona, \#coronavirus, \#Coronavirus, \\ 

& \#coronavirus2020, \#coronavirususa, \#outbreak, \#CoronavirusOutbreak, \#CoronaOutbreak\\

& \#coronavirusoutbreak, \#covid19outbreak, \#coronaapocalypse, \#cononavirusPandemic, \\

& \#nyccoronavirus, \#bayareacoronavirus, \#seattlecovid19, \#Floridacoronavirus, \#utahcovid19, \\

& \#ohiocoronavirus , \#coronavirusupdate,  \#californiacoronavirus \\
& \#pandemic  \#CoronaVirusUpdates and \#highriskcovid19 \\
\hline

Quarantine & \#QuarantineLife, \#quarantined, \#staysafestayhome, \#staytheFhome, \#staythefuckhome, \\

&   \#Quarantinelife,\#socialdistancing, \#SocialDistancing, \#SocialDistancingNow, \#Quaratineandchill, \#workfromhome, \\

&  \#homeoffice, \#selfquarantine, \#stayhome,  \#stayhomechallenge, \#quarantinecats, \\

& \#workingfromhome, \#togetherapart, \#flattenthecurve and \#flatteningthecurve\\

 \hline
 
 School Closures &\#schoolclosures, \#closenypublicschools, \#closenycschools, \#closenyschools, \#suny, \#cuny, \\

& \#homeschool, \#homeschooling, \#Homeschooling, \#homeschool2020, \#schoolsclosed, \#noschool, \\

& \#closetheschools, \#shutdownschools, \#closethepublicschools, \#schoolsclosing, \#CloseTheSchools \\

\hline
Panic Buying & \#panicbuying, \#PanicBuying, \#panicshopping, \#PanicShopping, \#panicbuyers, \#toiletpaper, \\

& \#notoiletpaper, \#handsanitizer, \#ToiletPaperPanic, \#toiletpapercrisis, \#toiletpapershortage, \\

& \#howmuchtoiletpaper, \#ToiletPaperApocalypse, \#coronashopping, \#WashYourHands, \\

& \#washyourhands and \#sanitizers,\\

\hline
 Lockdowns & \#lockdown, \#Shutdown, \#seattleshutdown, \#ShutItDown, \#shutdownnyc, \#lalockdown, \#sflockdown, \\
 
 &\#bayarealockdown, \#lockdownusa, \#californialockdown, \#vegasshutdown, \\
 
&  \#newjerseylockdown, \#barsclosed, \#barsshut, \#californiashutdown, \\

& \#newyorklockdown, \#illinoislockdown, \#shutdownflorida, \#floridalockdown, \#ohiolockdown, \\
& \#ShutDownMass, \#ShutDownNYC \#NYCLockdown and \#nycshutdown \\

\hline
 Frustration and Hope & \#askthemayor, \#coronavirusremedy, \#fuckcovid19, \#canceltherent, \#CancelRent,  \\
 & \#fightcorona, \#fightback,\#saveyourlife, \#COVIDIOTS, \#COVIDIDIOTS, \#DrFauci, \#coronafighters, \\
  & \#letsfightcorona, \#stopthespread, \#saveworkers, \#SaveTheDay \#savelives \#StopTheSpread, \\
  & \#StayHomeSaveLives, \#WhenThisIsAllOver and \#StaySafe \\

\hline
\end{tabular}
\label{tab_hashtags}
\end{center}
\end{table*}

\begin{table*}[ht]
\caption{Example Tweets by Category}
\vspace{-2 mm}
\begin{center}
\begin{tabular}{|c|c|}
\hline
\textbf{Category} & \textbf{Example Tweets} \\
\hline
General COVID &  If this wasn't so f***ing deadly serious I'd be laughing... \#COVID19\\
 \hline
Quarantine &  Extroverts. I get it. You need human interaction to fuel your well being, \\
& but \#StayTheFHome and interact on social media  \\

 \hline
School Closures & If \@ NYCMayorsOffice \@ NYCMayor won't \#closenycpublicschools to protect students and their families, we  \\
& will \#sickout  \#CLOSENYCPUBLICSCHOOLS. Teachers are parents too. We all have family. Keep us all safe. \\

 \hline
Panic Buying & Stop hoarding toilet paper, you morons!   \#ToiletPaperPanic \\

\hline
 Lockdowns & NYC is going to get destroyed. I'm so depressed.\#NYCLockdown \\

 \hline
 Frustration and Hope & When are we going to \#CancelRent in this state? Hundreds of thousands are filing for unemployment and \\
 & can't pay rent. Sure, we can't be evicted, but what's preventing companies from coming after us after this is over? \\

  \hline
\end{tabular}
\label{tab_tweets}
\end{center}
\end{table*}

\subsection{Hashtag Categories}
We group the hashtags into six main categories, namely {\it 1)} General COVID, {\it 2)} Quarantine,  {\it 3)} School Closures, {\it 4)}  Panic Buying, {\it 5)} Lockdowns, and  {\it 6)} Frustration and Hope  to quantitatively and qualitatively understand the chain of events. We collect data on per day basis for the different hashtags as and when they become trending. Tables \ref{tab_total_tweets} and  \ref{tab_hashtags} show the number of tweets in each category and the grouping of the hashtags by category. We observe that the total number of tweets as grouped by hashtags is 664,476, which is higher than the total number of tweets. This is because tweets can contain multiple hashtags and thus the same tweet can be  grouped into multiple categories. We present some example tweets in Table \ref{tab_tweets} to illustrate the types of communications occurring on Twitter during this period.

\noindent {\bf 1. General COVID:}  In this category, we group hashtags related to COVID related messages as it is the most discussed topic in conversations. This grouping is done by accumulating hashtags related to COVID-19. 


\noindent {\bf 2. Quarantine:} Calls for social distancing and quarantines  flooded Twitter during this outbreak. Communications  centered around quarantines, working from home and flattening the curve to slow the spread of the virus. 

\noindent {\bf 3. School Closures:} In this category, we  collect data related to school closures.   Before states decided to close schools, users on Twitter  demanded the government to shut down public schools and universities. We collect data from a number of  hashtags centered around this call for action.


\noindent {\bf 4. Panic Buying:} The spread of the virus also resulted in panic buying  and hoarding. People  rushed to shopping marts and there was a huge panic buying of sanitizers and toilet paper. This panic buying resulted in severe shortage of toilet papers around the middle of March, an issue that remained unresolved till the first week of April.


\noindent {\bf 5. Lockdowns:}  With COVID-19 spreading unabated, lock downs of public stores, bars, restaurants, and cities began in many states. This resulted in a surge in tweets related to lock downs.



\noindent {\bf 6. Frustration and Hope:}  Emotions  ran high during these times with people expressing anger and resentment towards those not abiding by social distancing and quarantine rules. Alongside, people also rallied to support workers working hard to keep essential services running. With the beginning of April approaching, many people started to  worry about their next month's rent. 


\subsection {Gaps in Data Collection} Due to the data collection limits imposed by Twitter, we are able to only collect and analyze a portion of the tweets. Though we started collecting data as quickly as we conceived of this project, we were unable to collect data during the first week of March. Though we ran our script to collect data as far back as March 8, because of the way Twitter provides data, we  obtained limited number of tweets from March 8 to March 13. Additionally, due to the rapidly evolving situation, it is likely that we have inadvertently missed some important hashtags, despite our best efforts. As is the case with most studies based on Twitter data, we also acknowledge the presence of bias in data collection \cite{morstatter:2017discovering}. Having said that, the goal of this study is to provide a panoramic summarized view of the impact of the pandemic on people's lives and aggregate public opinion as expressed by them. Due to the nature of this study, we  are confident that the results presented here help in appreciating the sequence of events that transpired and better prepare ourselves from a possible second outbreak of COVID-19 or another pandemic.

\section{Data Analysis}
\label{sec:experiment}

In this section, we present  observations and results based on our preliminary analysis of the tweets. We study the popularity of individual hashtags and investigate how the number of tweets in particular hashtag groups evolve over time. We also explore the term frequency for each hashtag group to understand the main points of discussion. Our analysis summarizes the critical public responses surrounding COVID-19 and paves the way for more insightful and fine-grained linguistic analysis in the future.

\subsection{Temporal Evolution of Hashtag Groups}
\begin{figure}
    \centering
  \subfloat[Number of tweets for the different hashtags]{%
       \includegraphics[scale=0.16]{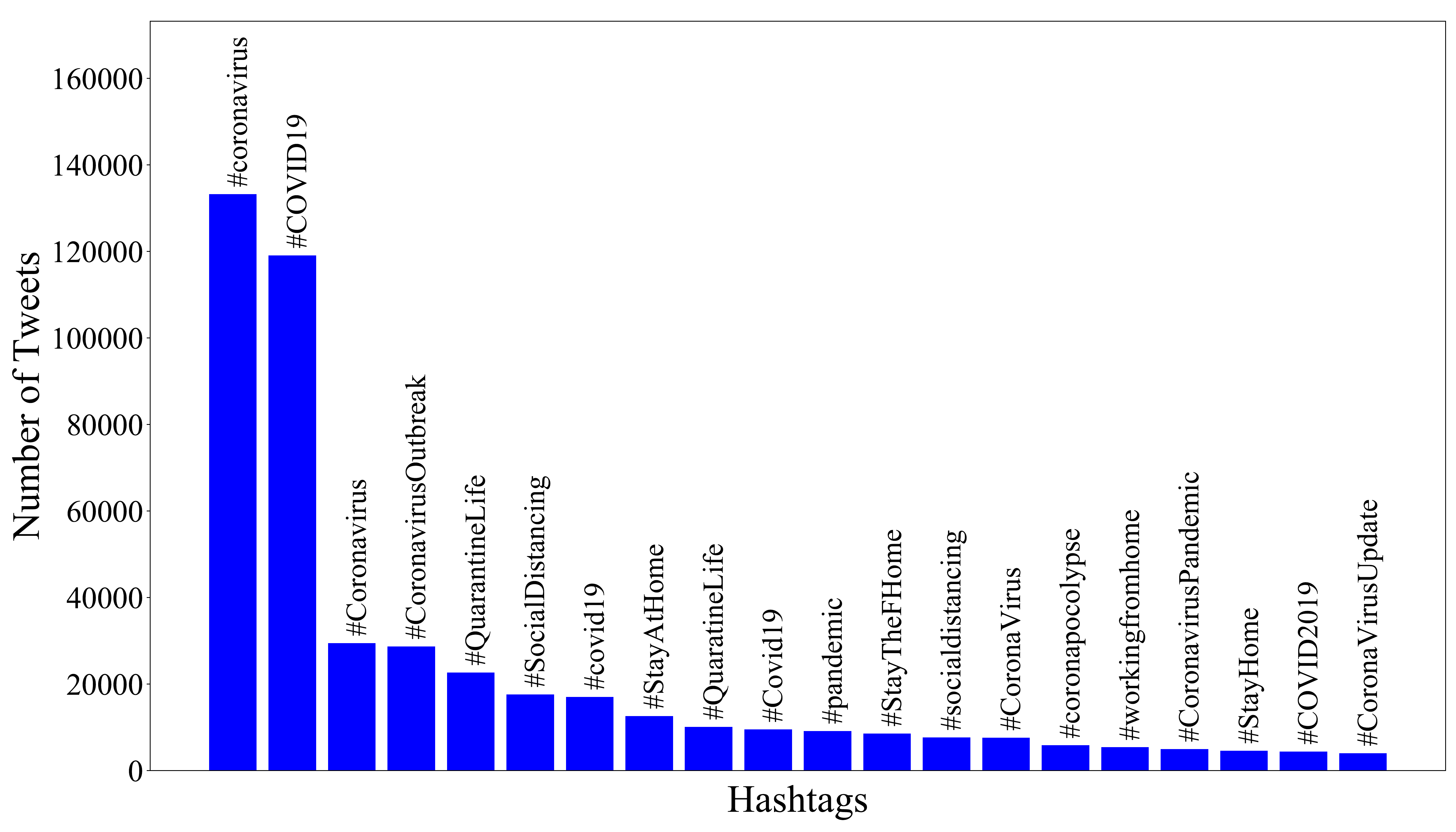}
       \label{fig:hastags_top}}
       \vspace{1mm}
  \subfloat[Most trending hashtags by day]{%
       \includegraphics[scale=0.16]{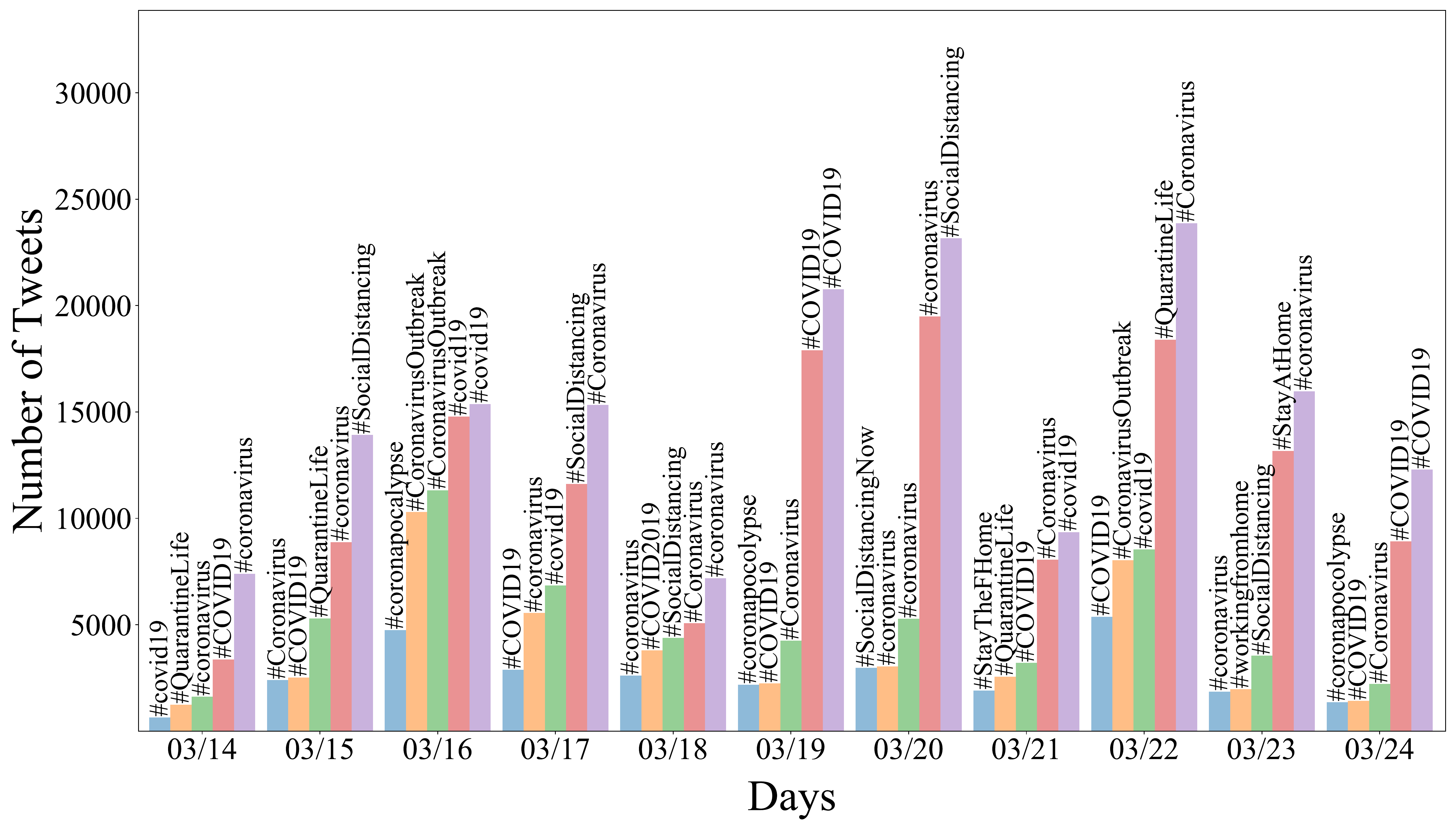}
       \label{fig:hastags_trending}}
       \vspace{1mm}
	\caption{Popularity of different hashtags} 
  \label{fig:hastags} 
    \vspace{-4mm}
\end{figure}

Figure \ref{fig:hastags_top} shows the top 20 hashtags observed in our data. As expected, we see that hashtags corresponding directly to COVID or coronavirus are the most popular hashtags as most communications are centered around them. We observe that hashtags around social isolation, staying at home, and quarantining are also popular. Figure \ref{fig:hastags_trending} shows the most popular hashtags by date. Similar to Figure \ref{fig:hastags_top}, we observe that hashtags related directly to COVID and social distancing trend most on Twitter. The figures and the number of tweets  highlight how the pandemic gripped the United States with its rate of spread.

\begin{figure*}[!ht]
    \centering
  \subfloat[General COVID and Quarantine]{%
       \includegraphics[scale=0.15]{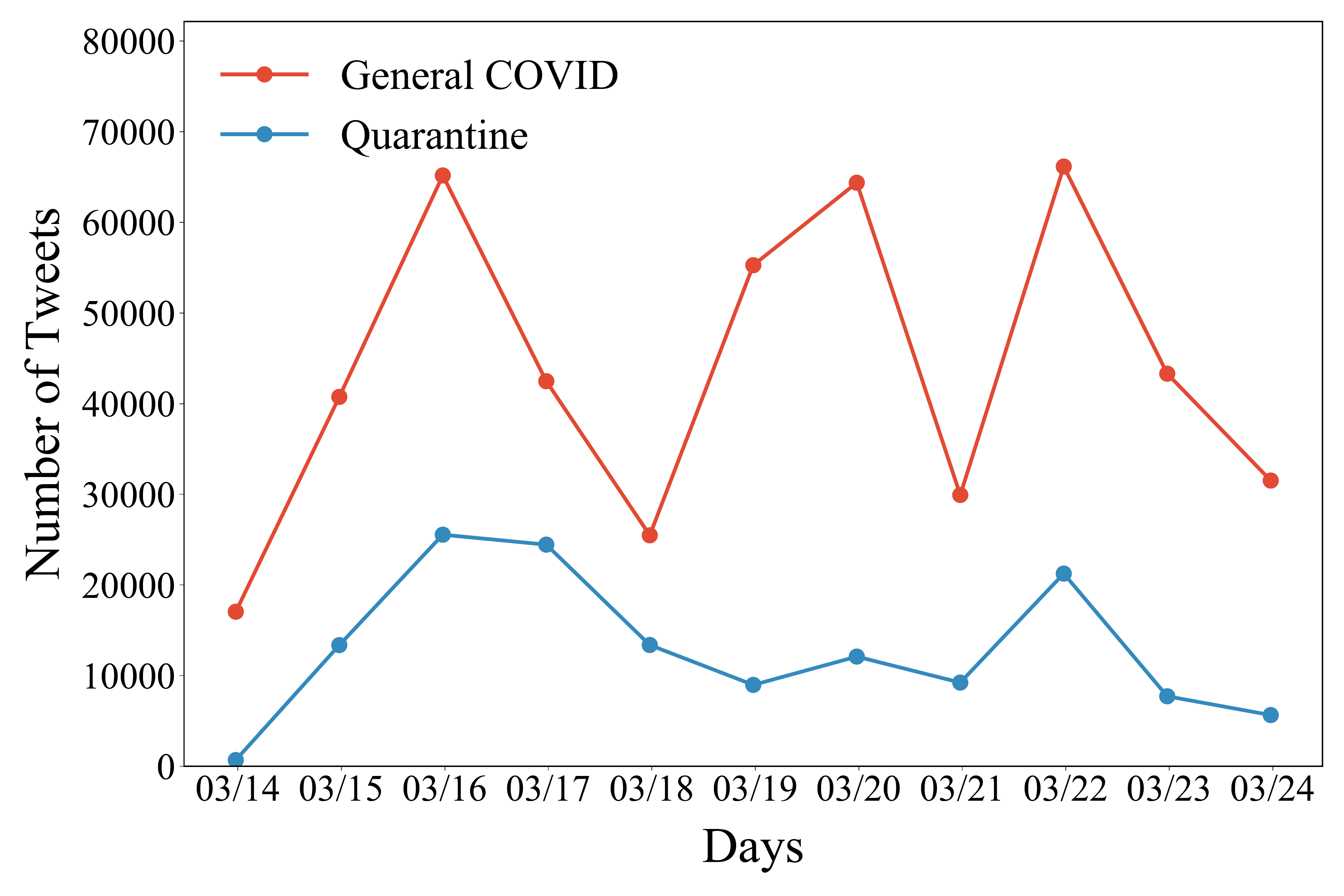}
       \label{fig:covid_quarantine}}
       \vspace{1mm}
  \subfloat[School Closures and Panic Buying]{%
       \includegraphics[scale=0.15]{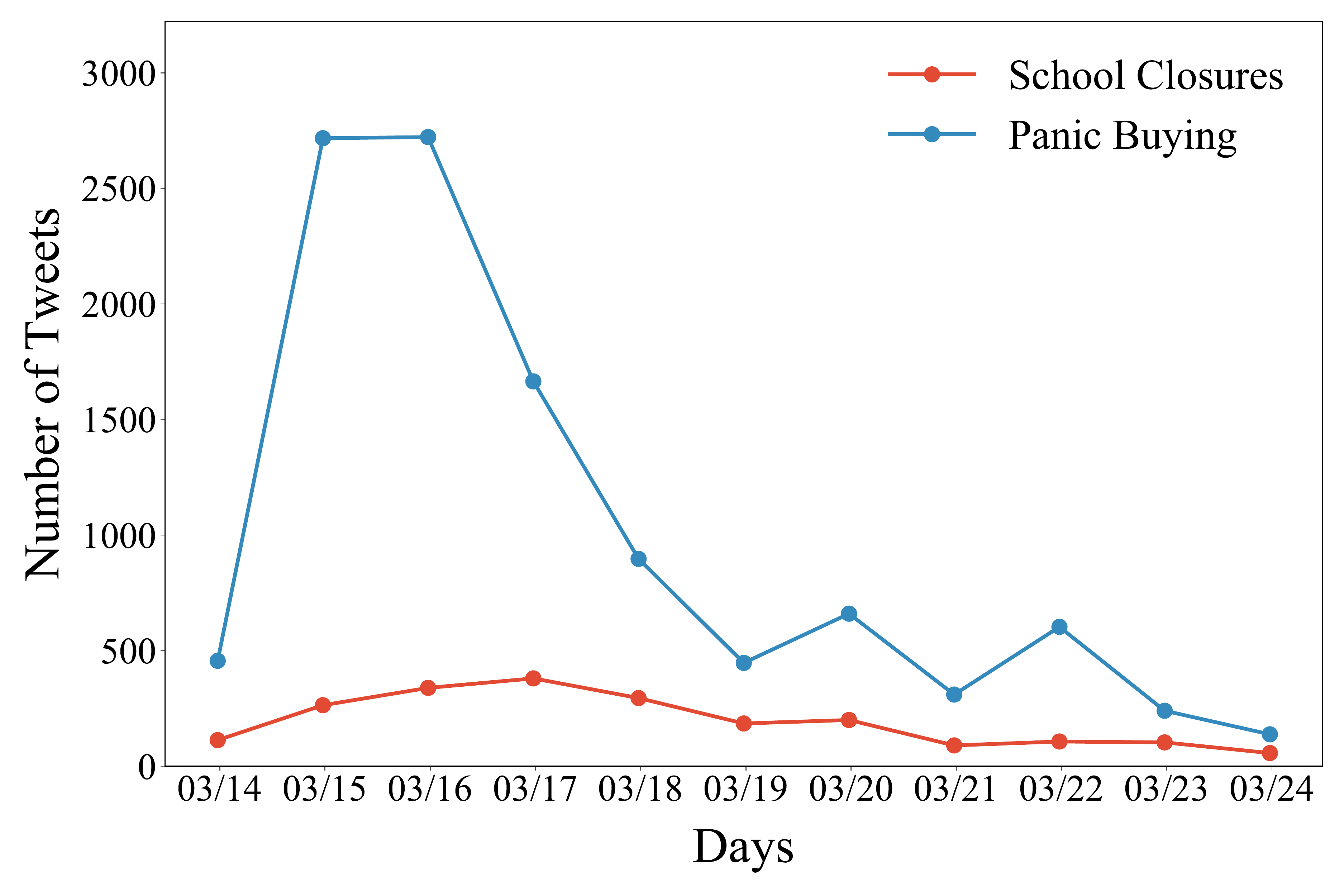}
       \label{fig:panic_school}}
       \vspace{1mm}
  \subfloat[Lockdowns and Frustration and Hope]{%
       \includegraphics[scale=0.15]{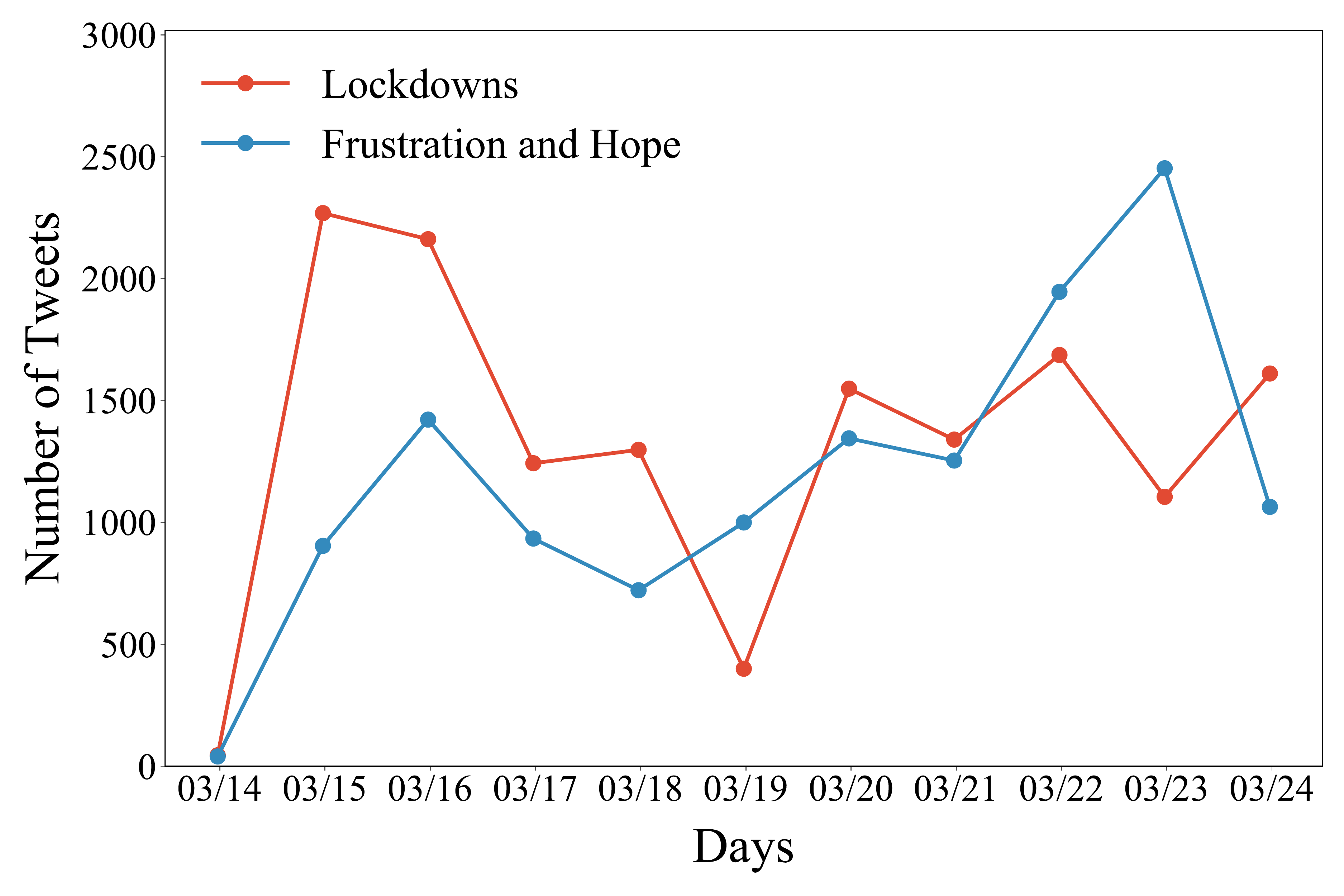}
       \label{fig:lockdown_frustation}}
	\caption{Temporal Evolution of Tweets in Hashtag Groups} 
  \label{fig:mae} 
    \vspace{-4mm}
\end{figure*}

We investigate the evolution of the number of tweets in various hashtag groups over time. To calculate the number of tweets in each hashtag group, we count the number of mentions of hashtags in that group across all the tweets. If the tweet contains more than one hashtag, it is counted as part of all the hashtags mentioned in it. As the number of tweets for hashtag groups vary significantly, we plot the groups that have similar number of tweets together. Similar to Figure \ref{fig:hastags}, we observe from Figure \ref{fig:covid_quarantine} that the total number of tweets in the General COVID and Quarantine categories are relatively high throughout the time period of the study.

Interestingly, from Figure \ref{fig:panic_school}, we observe that panic buying and calls for school closures peak around the middle of March and then decrease as school closures  and rationing of many essential goods such as toilet paper, cereal, and milk take effect. From Figure \ref{fig:lockdown_frustation}, we see that calls for lock downs related to schools, bars, and cities  peak in the middle of March. With the virus spreading unabated,  we observe intense calls for lock downs of cities and entire states around the beginning of the fourth week of March, resulting in an increased number of tweets in this category. With passage of time, we observe people increasingly expressing their frustration and distress in communications, while some hashtags attempt to inject a more positive outlook. 

\begin{figure*}[!ht]
    \centering
  \subfloat[General COVID word frequency]{%
       \includegraphics[scale=0.28]{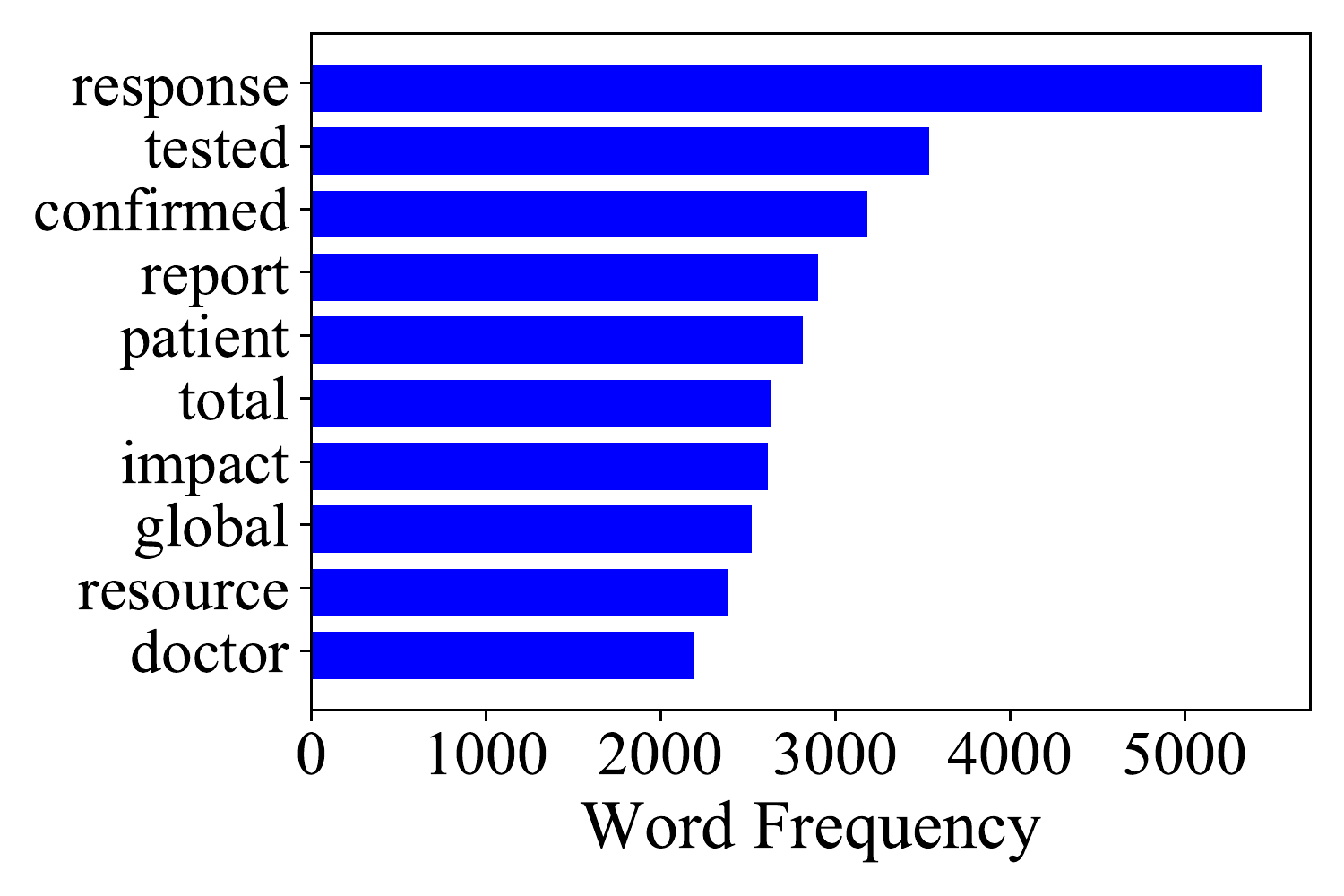}
       \label{fig:covid_word}}
       \vspace{1mm}
  \subfloat[School Closures word frequency]{%
       \includegraphics[scale=0.28]{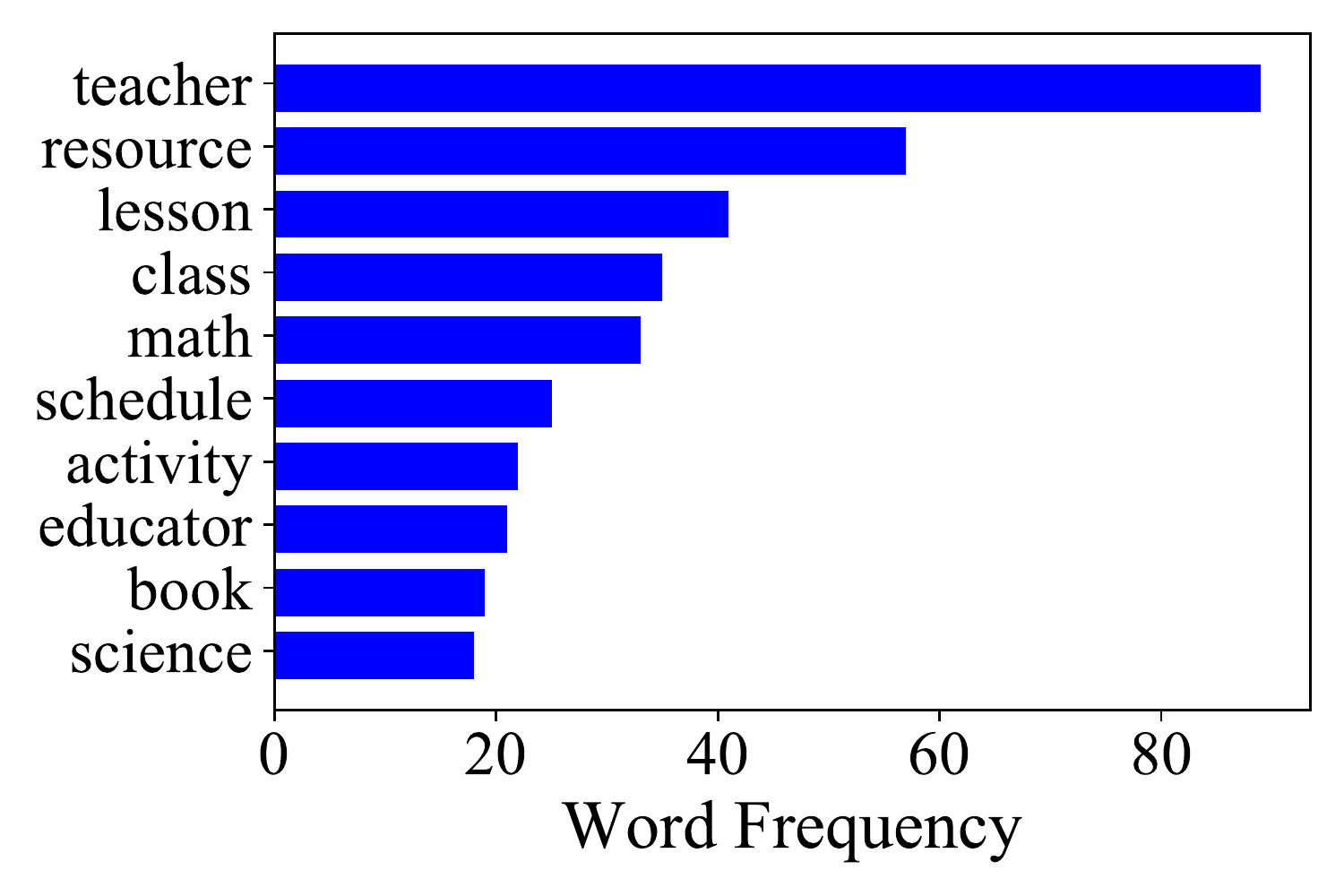}
       \label{fig:school_word}}
       \vspace{1mm}
  \subfloat[Panic Buying word frequency]{%
       \includegraphics[scale=0.28]{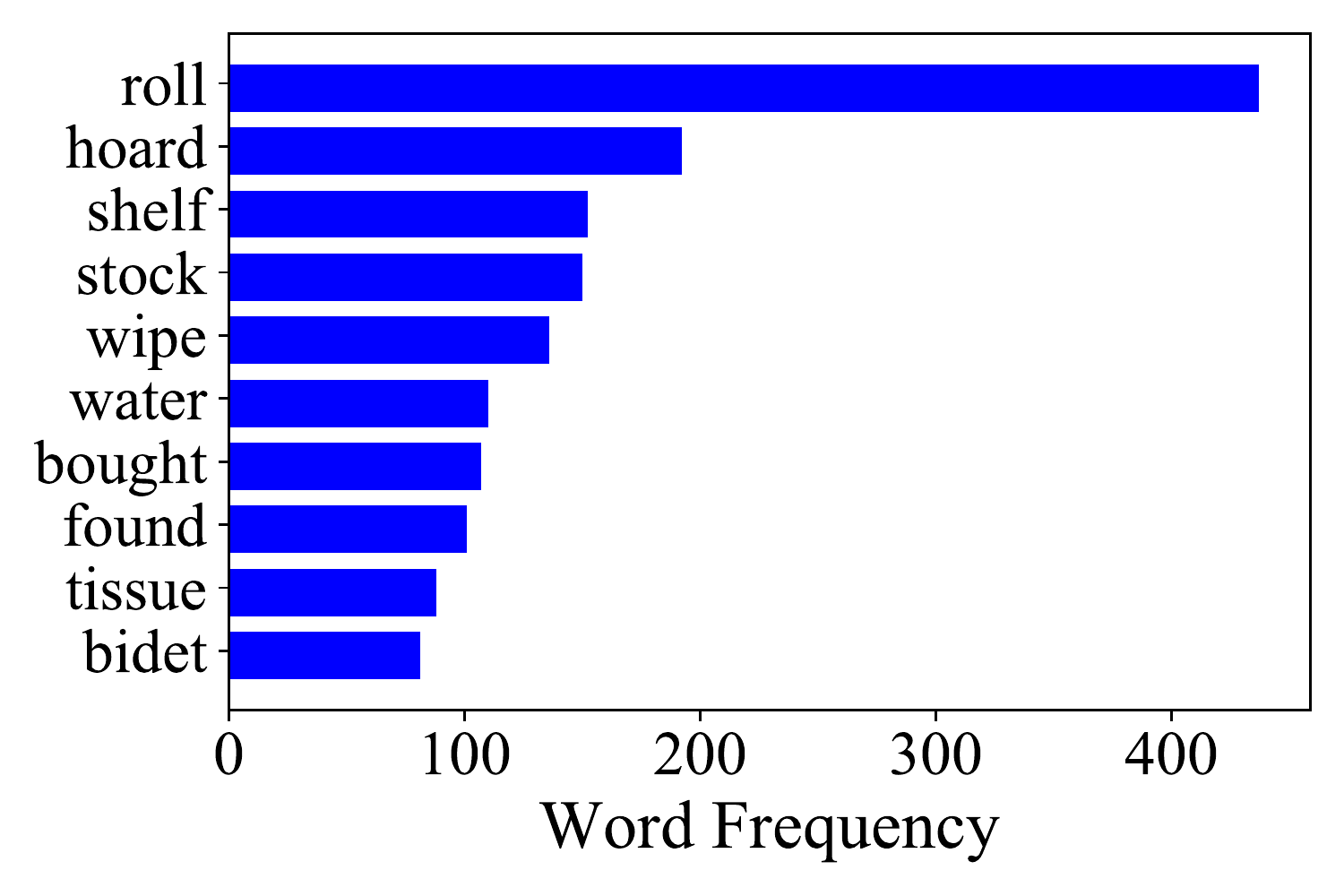}
       \label{fig:panic_word}}
       \vspace{1mm}
  \subfloat[Lockdowns word frequency]{%
       \includegraphics[scale=0.28]{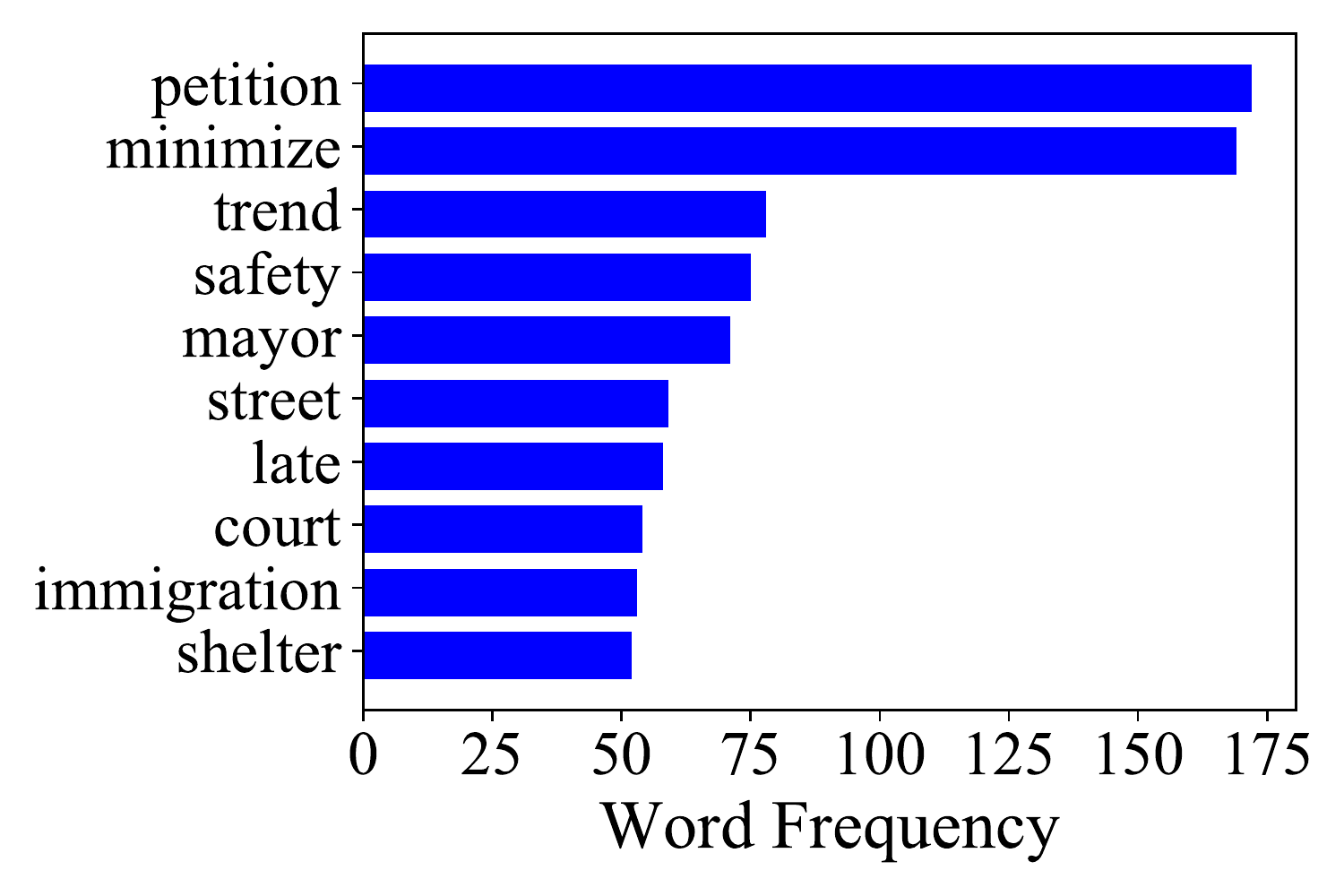}
       \label{fig:lockdown_word}}
	\caption{Word Frequencies} 
  \label{fig:words} 
    \vspace{-4mm}
\end{figure*}

\subsection{Linguistic Word-Usage Analysis}

In this section, we present results from a linguistic word usage analysis across the different hashtag groups. First, we identify and present the most commonly used words across all the hashtags. To construct the first group of common words across all hashtags, we remove the words that are same or similar to the hashtags mentioned in Table \ref{tab_hashtags} as those words are redundant and tend to also be high in frequency. We also remove the names of places and governors such as New York, Massachusetts and  Andrew Cuomo. After filtering out these words, we then rank the words based on their occurrence in multiple groups and their combined frequency across all the groups. We observe words such as \textit{family}, \textit{health}, \textit{death}, \textit{life}, \textit{work}, \textit{help},  \textit{thank}, \textit{need}, \textit{time}, \textit{love}, \textit{crisis}. In Table \ref{table:examples}, we present some notable example tweets containing the common words. While one may think that \textit{health} refers to the virus-related health issues, we notice that many people also refer to \textit{mental health} in their tweets as a possible consequence to social distancing and anxiety caused by the virus. We also observe the usage of words such as \textit{death} and \textit{crisis} to indicate the seriousness of the situation. Supporting workers and showing gratitude toward them is another common tweet pattern that is worth mentioning. We plan to use these observations to guide our future exploration and fine-grained analysis.

\begin{table}[ht]
\caption{Number of Tweets by Category}
\vspace{-2 mm}
\begin{tabular}{|M{7cm}|}
\hline
\textbf{Example of Tweets} \\
\hline
It’s more important than ever to prioritize your mental \textbf{health}. You are not in this
alone.\\
 \hline
Has $13^{th}$ century returned back to $20^{th}$ century? Black \textbf{Death}. We must act very fast. \\
\hline

First, we take care of the \textbf{health} care and emergency \textbf{workers}. Then, we take care of whoever is in charge of keeping Netflix and Hulu running or it’s going to get ugly \#distancesocializing \#coronavirus 
\\
 \hline
\end{tabular}
\label{table:examples}
\end{table}

Second, we present the most semantically meaningful and uniquely identifying words in each hashtag group. To do this, we remove the common words calculated in the above step from each group. From the obtained list of words after the filtering, we then select the top 10 words. Due to lack of space, we only present results for four hashtag groups. Figure \ref{fig:words} gives us the uniquely identifying and semantically meaningful words in each hashtag group. In the General COVID group, we find words such as impact, response, resource, doctor. Similarly, for School Closures, we find words such as teacher, schedule, educator, book, class. The Panic Buying top words mostly resonates the shortages experienced by people during this time such as roll and tissue (referring to toilet paper), hoard, bidet (as an alternative to toilet paper), wipe, and water. Top words in the Lockdowns group include immigration, shelter, safety, court, and petition, signifying the different issues surrounding lockdown.   


\section{Related Work}
\label{sec:related}

In this section, we  outline existing research related to modeling and analyzing twitter and web data to understand social, political, psychological and economic impacts of a variety of different events. Due to the recent nature of the outbreak, as far as we are aware there are no published research results related to COVID-19. Due to the space limitations, we only discuss social-media analysis work that are closely related to our work. Twitter has been used to study political events and related stance \cite{JG_coling_2016,le2017bumps}, human trafficking \cite{tomkins2018impact}, and public health \cite{perez2019using,Dredze:IS12,Balani:DCM15,Choudhury:AAAI13,McIver:jmir15}. Several work perform fine-grained linguistic analysis on social media data \cite{zhang2018fine,tomkins2018socio,raisi2018weakly}.


\section{Discussion and Concluding Remarks}
\label{sec:conclusion}
In this paper, we studied Twitter communications in the United States during the early days of the COVID-19 outbreak.  As the disease continued to spread, we observed that calls for closures of schools, bars, cities and entire states as well as social distancing and quarantining quickly gained fervor. Alongside, we observed an increase in panic buying and lack of availability of essential items, in particular toilet paper. We also conducted a linguistic word-usage analysis and distinguished between words that are common to multiple hashtag groups and words that uniquely and semantically identify each hashtag group. In addition to words that represent the group, our analysis unearths many words related to emotion in the tweets. Also, our qualitative analysis reveals that the words are used in multiple different contexts, which is worth delving further into.

The research presented in this paper is preliminary and its primary aim is to quantitatively outline the socio-economic distress already caused by COVID-19 so that we as a society can learn from this experience and be better prepared if COVID-19 (or maybe another pandemic) were to (re)emerge in the future.  At the time of writing this paper, the infection spread has still not reached its peak in the United States. Therefore, we plan to keep collecting data to understand and investigate the socio-economic and political impact of COVID-19. We plan to expand on our current research by performing topic modeling and expanding our linguistic analysis to unearth the main topics being discussed in the tweets. We also plan to conduct sentiment analysis to understand the extent of positive and negative sentiments in the tweets.

\bibliographystyle{ACM-Reference-Format}
\bibliography{references}

\end{document}